\documentclass[sn-nature]{sn-jnl}

\usepackage[utf8]{inputenc}
\usepackage[usenames,dvipsnames]{color}
\usepackage{amsmath}
\usepackage{amssymb}
\usepackage{mathptmx}
\usepackage{CJKutf8}

\usepackage{multirow}
\usepackage{array}
\usepackage{rotating}
\usepackage{booktabs}%

\title{\textbf{The dominance of turbulence over magnetism in the formation of massive star cluster seeds}}

\begin{document}

\begin{CJK*}{UTF8}{gbsn} 
\author*[1,2,3]{Junhao Liu (刘峻豪)}\email{liujunhao@nju.edu.cn}
\affil*[1]{School of Astronomy and Space Science, Nanjing University, 163 Xianlin Avenue, Nanjing, Jiangsu 210023, People's Republic of China}
\affil[2]{Key Laboratory of Modern Astronomy and Astrophysics (Nanjing University), Ministry of Education, Nanjing, Jiangsu 210023, People's Republic of China}
\affil[3]{National Astronomical Observatory of Japan, 2-21-1 Osawa, Mitaka, Tokyo 181-8588, Japan}

\author[4]{\fnm{Patricio} \sur{Sanhueza}} 
\affil[4]{Department of Astronomy, School of Science, The University of Tokyo, 7-3-1 Hongo, Bunkyo, Tokyo 113-0033, Japan}

\author[5,3]{Piyali Saha}
\affil[5]{Academia Sinica, Institute of Astronomy and Astrophysics, No.1, Sec. 4, Roosevelt Road, Taipei 10617, Taiwan, R.O.C.}

\author[6,4]{Kaho Morii}
\affil[6]{Center for Astrophysics | Harvard \& Smithsonian, 60 Garden Street, Cambridge, MA 02138, USA}

%

\author[7,8]{Josep Miquel Girart}
\affil[7]{Institut de Ciencies de l'Espai (ICE, CSIC), Can Magrans s/n, 08193, Cerdanyola del Vallès, Catalonia, Spain}
\affil[8]{Institut d'Estudis Espacials de Catalunya (IEEC), 08034, Barcelona, Catalonia, Spain}

\author[6]{Qizhou Zhang}

\author[3]{Fumitaka Nakamura}

\author[9,10]{Paulo C. Cort\'es}
\affil[9]{Joint ALMA Observatory, Alonso de C\'ordova 3107, Vitacura, Santiago, Chile}
\affil[10]{National Radio Astronomy Observatory, 520 Edgemont Road, Charlottesville, VA 22903, USA}

\author[11,12]{Valeska Valdivia}
\affil[11]{Department of Physics, Graduate School of Science, Nagoya University, Furo-cho, Chikusa-ku, Nagoya 464-8602, Japan}
\affil[12]{Liant, 5 Av. Jean Jaur\`es, 34600, B\'edarieux, France}

\author[13]{Beno\^it Commer\c{c}on}
\affil[13]{Univ. Lyon, Ens de Lyon, Univ. Lyon 1, CNRS, Centre de Recherche Astrophysique de Lyon UMR5574, 69007, Lyon, France}

\author[5]{Patrick M. Koch}

\author[14]{Kate Pattle}
\affil[14]{Department of Physics and Astronomy, University College London, Gower Street, London WC1E 6BT, United Kingdom}

\author[15,16]{Xing Lu}
\affil[15]{Shanghai Astronomical Observatory, Chinese Academy of Sciences, 80 Nandan Road, Shanghai 200030, People’s Republic of China}
\affil[16]{100101 Key Laboratory of Radio Astronomy and Technology (Chinese Academy of Sciences), A20 Datun Road, Chaoyang District, Beijing, 100101, People’s Republic of China}

\author[14]{Janik Karoly}

\author[17]{Manuel Fern\'andez-L\'opez}
\affil[17]{Instituto Argentino de Radioastronomía (CCT- La Plata, CONICET, CICPBA, UNLP), C.C. No. 5, 1894, Villa Elisa, Buenos Aires, Argentina}

\author[18]{Ian W. Stephens}
\affil[18]{Department of Earth, Environment, and Physics, Worcester State University, Worcester, MA 01602, USA}

\author[19]{Huei-Ru Vivien Chen}
\affil[19]{Institute of Astronomy and Department of Physics, National Tsing Hua University, Hsinchu 300044, Taiwan, R.O.C.}

\author[20]{Chi-Yan Law}
\affil[20]{INAF-Osservatorio Astrofisico di Arcetri, Largo E. Fermi 5, I-50125 Firenze, Italy}

\author[1,2]{Keping Qiu}

\author[1,2]{Shanghuo Li}

\author[21]{Henrik Beuther}
\affil[21]{Max Planck Institute for Astronomy, K\"onigstuhl 17, 69117 Heidelberg, Germany}

\author[22]{Eun Jung Chung}
\affil[22]{Korea Astronomy and Space Science Institute (KASI), 776 Daedeokdae-ro, Yuseong-gu, Daejeon 34055, Republic of Korea}

\author[23]{Jia-Wei Wang}
\affil[23]{East Asian Observatory, 660 N. A'oh\={o}k\={u} Place, University Park, Hilo, HI 96720, USA}

\author[24,3]{Fernando A. Olguin}
\affil[24]{Yukawa Institute for Theoretical Physics, Kyoto University, Kyoto, 606-8502, Japan}

\author[3]{Yu Cheng}

\author[25,26]{Jihye Hwang}
\affil[25]{Institute for Advanced Study, Kyushu University, Japan}
\affil[26]{Department of Earth and Planetary Sciences, Faculty of Science, Kyushu University, Nishi-ku, Fukuoka 819-0395, Japan}

\author[27]{Sandhyarani Panigrahy}
\affil[27]{Department of Physics, Indian Institute of Science Education and Research (IISER) Tirupati, Yerpedu, Tirupati - 517619, Andhra Pradesh, India}

\author[27]{Chakali Eswaraiah}

\author[20]{Maria T.\ Beltr\'an}

\author[4]{Qiuyi Luo}

\author[22]{Spandan Choudhury}

\author[22]{Ji-hyun Kang}

\author[15]{Wenyu Jiao}

\author[28]{Luis A. Zapata}
\affil[28]{Instituto de Radioastronom\'\i a y Astrof\'\i sica, Universidad Nacional Aut\'onoma de M\'exico, P.O. Box 3-72, 58090, Morelia, Michoac\'an, M\'exico}

\author[22]{A. -Ran Lyo}



\abstract{High-mass stars form in protoclusters, where gravo-magnetic processes shape collapsing clouds and clumps to be elongated preferentially perpendicular to magnetic (B) fields. Yet it remains unclear whether gravo-magnetic processes still govern the formation of smaller-scale condensations in massive-star-forming protoclusters, which are crucial for understanding the stellar initial mass function and multiplicity. 
Here we report the first statistical evidence that the condensation elongations are preferentially aligned with local B fields, based on high-resolution data from the largest dust polarization survey toward 30 massive star-forming regions with the Atacama Large Millimeter/submillimeter Array (ALMA).  Our clustered massive star formation simulations reveal that this more parallel alignment is exclusively observed in models where initial turbulence dominates B fields. In contrast, models with initial B fields dominating turbulence distinctly exhibit a more perpendicular alignment. 
The comparison between observations and simulations suggests that turbulence could play a more important role than B fields in the formation of condensations in the context of clustered massive star formation, contradicting the prediction of classical magnetically regulated models. Moreover, we find a possibly turbulence-induced preferential misalignment between the B field and rotation axis of condensations, which may potentially reduce the magnetic braking efficiency and facilitate the formation of large protostellar disks. Our findings indicate that turbulence could be critical in determining the initial stellar properties.
}


\maketitle
\end{CJK*}
\newpage
\section*{Main}

Nearly all high-mass stars ($M_{*}>8M_{\odot}$) form in clusters \cite{2025ARA&A..63....1B}, which originate from the conversion of gas into stars within $\sim$10 pc massive-star-forming interstellar clouds. This process begins with the collapse of molecular gas due to self-gravity, leading to hierarchical fragmentation into compact gas structures of $\sim$1 pc clumps, $\sim$0.1 pc cores, and $\sim$0.01 pc condensations in protoclusters. Condensations are the immediate parental structures of protostellar disks. Once gravity dominates over other forces, an individual condensation may collapse to form a single star or a close multiple system \cite{2025ARA&A..63....1B}. Unveiling the formation mechanism of condensations is thus crucial for understanding the origin and properties of stellar clusters in the high-mass regime \cite{2004MNRAS.349..735B}.

Gravo-magnetic processes are thought to play an important role in shaping the density distribution of collapsing clouds and their substructures \cite{2014ApJ...792..116Z, 2015Natur.520..518L}.  This is because magnetic (B) fields are energetically more significant than turbulence and gravity at cloud scales \cite{2016AA...586A.138P}, and gravity overcomes both B fields and turbulence as density increases within molecular clouds \cite{2013ApJ...779..185K, 2022ApJ...925...30L}. In such a scenario, gas collapse is expected to proceed more efficiently along field lines than in the perpendicular direction, resulting in structures elongated perpendicular to the field \cite{1976ApJ...207..141M, 2008ApJ...687..354N}. Such more perpendicular alignment has been widely observed at $>$0.1 pc scales \cite{2009MNRAS.399.1681T, 2014ApJ...792..116Z, 2015Natur.520..518L}. 

At smaller scales, however, the physical condition and mechanism governing the formation of condensations in massive-star-forming protoclusters are far less clear \cite{2025ARA&A..63....1B}. In the lack of statistical observational constraints on the condensation-scale B property, and without systematic comparisons between simulations and observations, it remains unresolved whether gravo-magnetic processes continue to be the dominant mechanism shaping the density structure of condensations in protoclusters \cite{2024A&A...686A.281B, 2025A&A...701A.217L}, or B fields are relatively less important in such processes \cite{2020ApJ...904..168B, 2025MNRAS.539.2307K}. 

\begin{figure}[htb]
\centering
\includegraphics[width=0.95\textwidth]{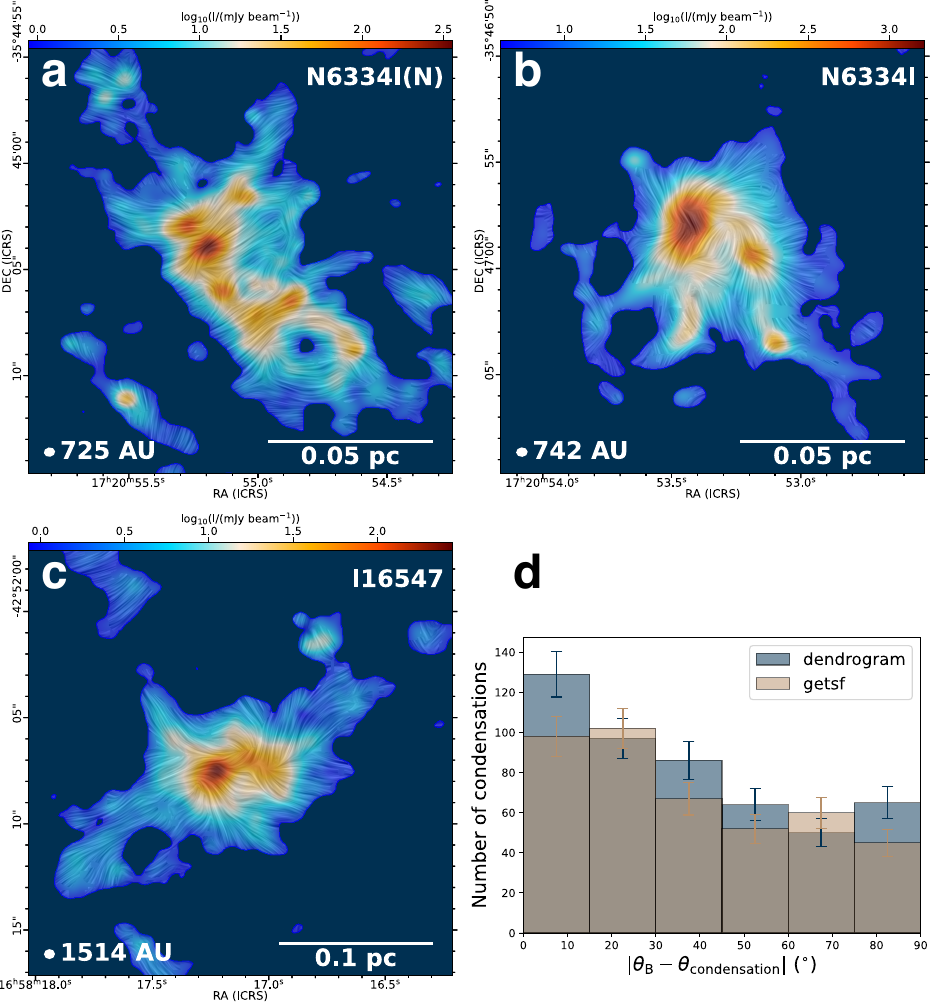}
\caption{\textbf{Properties of observed massive protocluster systems.} \textbf{a--c}, Examples of ALMA dust polarization observations  \cite{2021ApJ...923..204C, 2024ApJ...972..115C, 2024ApJ...974..257Z}. The background color shows the 1.3 mm dust intensity. Overlaid line patterns, generated using the line integral convolution method, indicate the B field orientation where $PI/\sigma_{PI}>2$. The synthesized beam is shown as a white ellipse in the lower left corner of each panel. A scale bar is shown in the lower right corner of each panel. \textbf{d}, Histograms (with Poisson errorbars) of the angular difference between B field orientation and condensation elongation across all observed regions. Different colors represent results from different source identification algorithms.
} \label{fig:obs}
\end{figure}

To probe this, we use the alignment between B field orientations and condensation elongations as a diagnostic \cite{2013MNRAS.436.3707L, 2015Natur.520..518L, 2014ApJ...792..116Z} of the outcome from different condensation formation scenarios. Thanks to the sensitivity of the Atacama Large Millimeter/submillimeter Array (ALMA), we have obtained a large sample of small-scale B field structures at $\sim$500-2000 au resolution through the Magnetic fields in Massive star-forming Regions (MagMaR) survey \cite{2021ApJ...915L..10S, 2021ApJ...923..204C, 2021ApJ...913...29F, 2024ApJ...972..115C, 2024ApJ...972L...6S, 2024ApJ...974..257Z, 2025ApJ...980...87S}. The MagMaR survey maps 250 GHz dust polarization across 30 massive star-forming regions, making it the largest ALMA dust polarization survey in the regime of high-mass star formation. Figs. \ref{fig:obs}a-c show representative examples of the B field orientations revealed by ALMA overlaid on dust continuum maps. We have identified the condensations from the dust intensity maps with commonly-used source identification algorithms \citep[{\tt astrodendro} and {\tt getsf};][]{2008ApJ...679.1338R, 2021A&A...649A..89M}, and measured the elongation ($\theta_{\mathrm{condensation}}$) and average B field orientation ($\theta_{\mathrm{B}}$) of each condensation (see Methods and Extended Data Fig. 1). The median Full Width at Half Maximum (FWHM) of the identified condensations is 957 au for {\tt astrodendro} and 1389 au for {\tt getsf} (i.e., with diameter $\sim$0.01 pc). To avoid potential emission contamination, we exclude a small portion of condensations ($\sim$2\%) that overlap with HII regions from our analysis. 

The comparison between condensation elongations and local B fields across all observed regions reveals a clear trend: the two orientations are preferentially more parallel (Fig. \ref{fig:obs}d). Note that uncertainties arising from source identification parameters, projection effects, condensation aspect ratios, beam smoothing effects, and dispersion of B field orientations do not significantly change the general alignment trend (see Supplementary Information). A Kolmogorov–Smirnov (K–S) test yields a low probability (0.03) that the observed alignment distribution arises from a random population, allowing us to reject the null hypothesis. This observed more parallel alignment contrasts with the preferential more perpendicular alignment previously reported at larger scales in massive star-forming regions \cite{2015Natur.520..518L, 2014ApJ...792..116Z}. 

Comparing the statistical trends from observations with those from simulations can establish links between their underlying physical properties \cite{2016AA...586A.138P}. However, no numerical studies have yet systematically examined the alignment between condensations and B fields in different physical conditions. To address this, we analyze synthetic observations of 11 models of self-gravitating clustered massive star formation simulations with varying magnetic and turbulent levels (see Methods). The simulations are generated with three-dimensional (3D) magneto-hydrodynamic (MHD) and radiative transfer codes with a box size of 1-2 pc and a maximum resolution of 12.5-25 au. Two models are initially sub-Alfv\'{e}nic with B fields dominating turbulence, while the other nine models are initially super-Alfv\'{e}nic with turbulence dominating B fields. The synthetic dust polarization maps are convolved to a resolution of 1000 au and include Gaussian noise at levels matching observational thermal noise.

\begin{figure}[htb]
\centering
\includegraphics[width=0.95\textwidth]{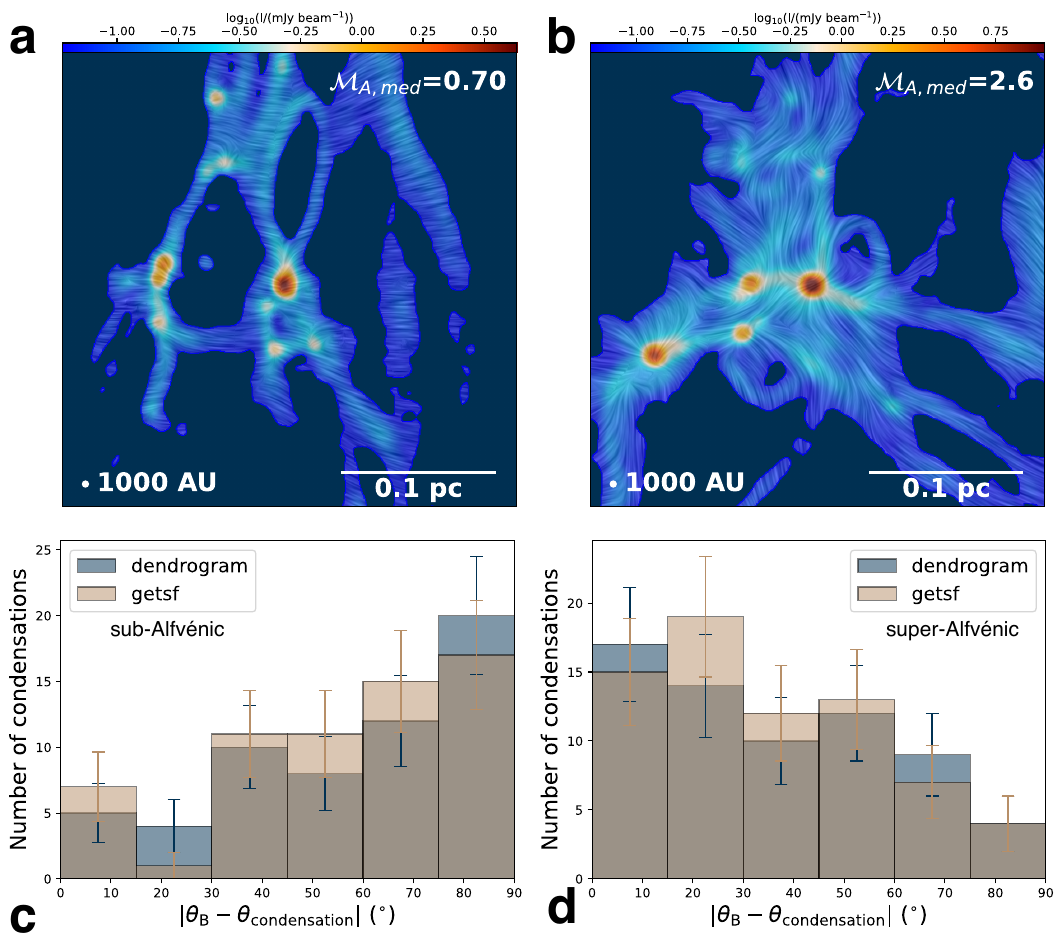}
\caption{\textbf{Properties of simulated massive protocluster systems.} \textbf{a--b}, Examples of synthetic observations for initially (a) sub-Alfv\'{e}nic (T10M3MU1) and (b) super-Alfv\'{e}nic (T10M6MU2) simulations. The initial median Alfv\'{e}nic Mach number $\mathcal{M}_{\mathrm{A,med}}$ (Table \ref{tab2}) is shown in the upper right corner of each panel. Maps are in the $xy$ plane and zoomed to the central 0.3 pc around the most massive protostar. The initial B field is along the $x$-axis (horizontal). The background color shows 1.3 mm dust intensity. Overlaid line patterns represent B field orientation via the line integral convolution method. \textbf{c--d}, Histogram examples of the angular difference between condensation elongation ($\theta_{\mathrm{condensation}}$) and average B field orientation ($\theta_{\mathrm{B}}$) in initially (c) sub-Alfv\'{e}nic (T10M3MU1) and (d) super-Alfv\'{e}nic (T10M6MU2) models, summed over 3 orthogonal planes.
} \label{fig:sim}
\end{figure}

Figs. \ref{fig:sim}a and b show examples of the B field orientation and dust intensity map from initially sub-Alfv\'{e}nic and super-Alfv\'{e}nic simulations, respectively. Condensations are identified in synthetic maps on three projected planes using {\tt astrodendro} and {\tt getsf}, following the same procedure as applied to observational data (see Methods and Extended Data Fig. 2). The sizes of the identified condensations are comparable to those observed. 

In initially sub-Alfv\'{e}nic models, condensation elongations exhibit a more perpendicular alignment with B fields (Fig. \ref{fig:sim}c), consistent with previous strong-field simulations at larger scales \cite{2008ApJ...687..354N}. This supports the idea that dynamically important B fields tend to produce high-density structures elongated more perpendicular to the field orientation (see Extended Data Fig. 3). 

In contrast, initially super-Alfv\'{e}nic models show a preferentially more parallel alignment between condensations and B fields (Fig. \ref{fig:sim}d). Such alignment trend is consistent with the scenario that the condensed density structures are preferentially assembled at the converging position of gravo-turbulent flows. At those converging positions, shock compression could flatten the condensations, aligning their shortest axes with the flow direction of shock propagation \cite{2001ApJ...559.1005P, 2017A&A...603A..64S}. Simultaneously, the shocks can compress and amplify the B field components perpendicular to the shock propagation direction \cite{2001ApJ...559.1005P, 2017A&A...603A..64S}. Consequently, the converging flows could naturally produce a more parallel alignment between the B field orientation and the projected elongation direction of condensations \cite{2001ApJ...559.1005P, 2017A&A...603A..64S}. We provide a schematic view of this process in Extended Data Fig. 3.

\begin{figure}[htb]
\centering
\includegraphics[width=0.95\textwidth]{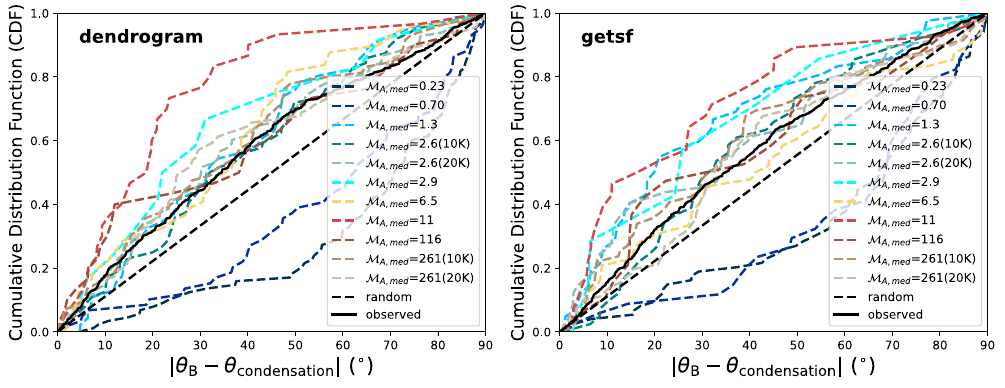}
\caption{\textbf{Cumulative distribution function of condensation-B alignment.} The solid black line shows the CDF across all observed regions. The dashed black line represents a random alignment distribution. Colored dashed lines represent CDFs from different simulation models.
} \label{fig:cdf}
\end{figure}

Fig. \ref{fig:cdf} compares the cumulative distribution function (CDF) of the observed condensation-B angular differences with those from different simulation models and a random distribution. The observed CDF closely matches those from initially super-Alfv\'{e}nic simulations, which exhibit more parallel alignment distributions. Our results suggest that turbulence could play a more important role than B fields in the formation of condensations in protoclusters. While the alignment differs clearly between initially sub- and super-Alfv\'{e}nic models, the degree of alignment shows no strong correlation with the initial Alfv\'{e}nic Mach number within each category. This implies that other factors may also influence the resulting alignment distribution, such as clump density, ratios between B fields/turbulence and gravity, or feedback after condensation formation. 

Note that filament fragmentation and gravitational flows along B field lines may also produce alignment between B fields and elongation of fragments \cite{2023ApJ...943...76M}. However, the observed separation lengths of condensations are much smaller than those predicted by filament fragmentation models \cite[][]{2024ApJ...974...95I}. Moreover, we find no strong correlation between the condensation elongation and the gravity direction (see Supplementary Information). This suggests that those alternative mechanisms are unlikely to be the dominant mechanism responsible for forming the clusters of condensations in our sample. 
 


\subsection*{Implications for clustered massive star formation} 

As stellar clusters originate from groups of condensations, the mass distribution of condensations (i.e., condensation mass function) is thought to be closely linked to the stellar initial mass function (IMF). Therefore, studying the formation of condensations is crucial for understanding the origin of the IMF and stellar multiplicity \cite{2022A&A...662A...8M}. In models of massive star formation that assume monolithic collapse, condensation masses are largely determined during fragmentation, with most of the stellar mass being accreted from the parental condensation  \cite{2014prpl.conf..149T}. B fields have long been considered significant in this scenario, as stronger fields are thought to reduce the number of fragments while increasing their masses \cite{2011ApJ...742L...9C}. 

Pilot observational studies have attempted to test the relation between B fields and small-scale fragmentation using Submillimeter Array (SMA) dust polarization data toward samples of 0.1 pc core-scale B fields: Ref. \cite{2021ApJ...912..159P} reported a tentative positive correlation between the normalized mass-to-flux ratio and the number of fragments, whereas ref. \cite{2024A&A...682A..81B} found no significant trend. However, due to the uncertainties on the estimated parameters, current SMA results neither confirm nor reject a causal relation between 0.1 pc core-scale B fields and small-scale fragmentation in protoclusters \cite{2024A&A...682A..81B}. 

Without the need to make various assumptions to estimate the B field strength, comparing the orientation of B fields with the elongation of density structures presents a more straightforward approach to directly probe the dynamical role of B fields in regulating gas morphology  \cite{2009MNRAS.399.1681T, 2015Natur.520..518L, 2014ApJ...792..116Z, 2020ApJ...895..142L}. The synergistic analysis of our ALMA observations and numerical simulations provides systematic evidence that overall turbulence could be more important than B fields in the formation of $\sim$0.01 pc condensations in massive-star-forming protoclusters. While B fields may still play a dominant role in the formation of a subset of condensations, such magnetically regulated condensation formation is unlikely the prevailing mode in clustered environments. Our findings suggest that the influence of B fields on the formation of small-scale compact structures in massive star-forming regions could be less significant than previously assumed in magnetized collapse models. Instead, turbulent flows could play a critical role in shaping the properties of condensations, which ultimately impact the characteristics of massive stellar clusters. 

It is important to emphasize that our results do not imply that B fields are unimportant in massive star formation overall. Observations show that massive star-forming clouds are trans-to-sub-Alfv\'{e}nic on large scales \cite{2023ApJ...945..160L, 2024ApJ...966..120L}. Strong large-scale B fields may aid in forming more massive molecular clumps, providing a larger mass reservoir for turbulent condensations to accrete from, ultimately facilitating the formation of more massive stars. The observed transition from sub-Alfv\'{e}nic clouds to super-Alfv\'{e}nic substructures with increasing density is likely driven by gravitational effects \cite{2019ApJ...871...98Z}. 



\subsection*{Misalignment between B fields and rotation axis}

Strong B fields can efficiently transport angular momentum away from the infalling gas, thereby braking rotation and suppressing the formation of rotationally-supported disks \cite{2003ApJ...599..363A}. This magnetic braking issue is particularly critical for the formation of massive-star-forming young stellar objects, which may require larger and more massive disks than their low-mass counterparts. Proposed solutions to the magnetic braking problem include non-ideal MHD effects, turbulent motions, and the misalignment between the B field and the rotation axis of the parental structure forming disks \cite{2018FrASS...5...39W}. Here, we focus on studying the B-rotation alignment with our observations and simulations. 

Observationally, the rotation axes of unresolved disks are often indirectly inferred from outflow axes. Comparisons between outflow axes and core-scale B field orientations in massive star formation regions have revealed no strong correlations \cite{2014ApJ...792..116Z}. However, since outflow directions can evolve over time, they may not reliably trace the instantaneous rotation axis at launch \cite{2014prpl.conf..451F}. Velocity gradients, in contrast, provide a more direct probe of rotational motions \cite{1997ApJ...475..211O}, but direct observational comparisons between velocity-gradient–inferred rotation axes and B field orientations remain scarce.

To probe the B-rotation alignment, we analyze the gas kinematics traced by CH$_3$CN $14_3 - 13_3$ ($E_\mathrm{u}/k_{\mathrm{B}} = 157$ K), a common kinematic tracer for hot molecular cores (HMCs) associated with massive star formation. Since CH$_3$CN is unlikely to be significantly contaminated by outflow emission \cite{2025A&A...699A..34M}, the observed velocity gradients should predominantly reflect rotational and/or infall motions. In the case of a flattened condensation, a velocity gradient along the major axis can be primarily explained in terms of rotation, while alignment with the minor axis is more likely due to infall motions \cite{1997ApJ...475..211O}. 

\begin{figure}[htb]
\centering
\includegraphics[width=0.95\textwidth]{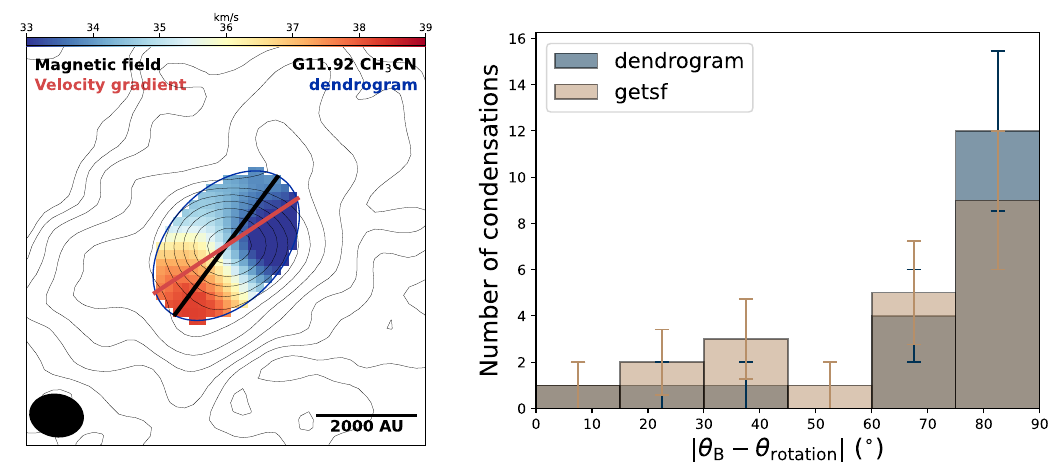}
\caption{\textbf{B-rotation alignment from observations.} Left: ALMA CH$_3$CN velocity centroid map of an example condensation. The blue ellipse outlines the structure identified by {\tt astrodendro}. The black and red lines indicate the average B field orientation and the velocity gradient direction, respectively. Contours represent dust continuum emission at levels of (3, 5, 8, 13, 21, 34, 55, 89, 144, 233, 377, 610, 987) $\times \sigma_I$. Right: Histogram of angular offsets between the average B field orientation and the inferred rotation axis of condensations. 
} \label{fig:obs_Brot}
\end{figure}

We classify rotation-like condensations as those with velocity gradients more aligned with the major axis (see Methods). The rotation axis  ($\theta_{\mathrm{rotation}}$) of each condensation is defined as the direction perpendicular to the velocity gradient, which is derived by fitting the CH$_3$CN velocity centroid map (see Methods). Fig. \ref{fig:obs_Brot} (left) presents the ALMA CH$_3$CN velocity centroid map of an example condensation. We further apply similar approaches to the simulation data and demonstrate that rotation axes derived from CH$_3$CN velocity gradients can reliably trace the intrinsic rotation axis of condensations (see Supplementary Information). 


The B-rotation alignment distribution across all observed regions is shown in Fig. \ref{fig:obs_Brot} (right). We find that B fields tend to be preferentially misaligned with the rotation axis. Note that the elongation of condensations in these cases is still preferentially more parallel to the local B fields. 

To further understand the origin of this misalignment, we perform a similar analysis using synthetic CH$_3$CN line data from numerical simulations (see Methods). Due to the limited number of condensations with well-defined rotation axes, we group results by sub-Alfv\'{e}nic and super-Alfv\'{e}nic models, summarizing the statistics from each category. 

\begin{figure}[htb]
\centering
\includegraphics[width=0.95\textwidth]{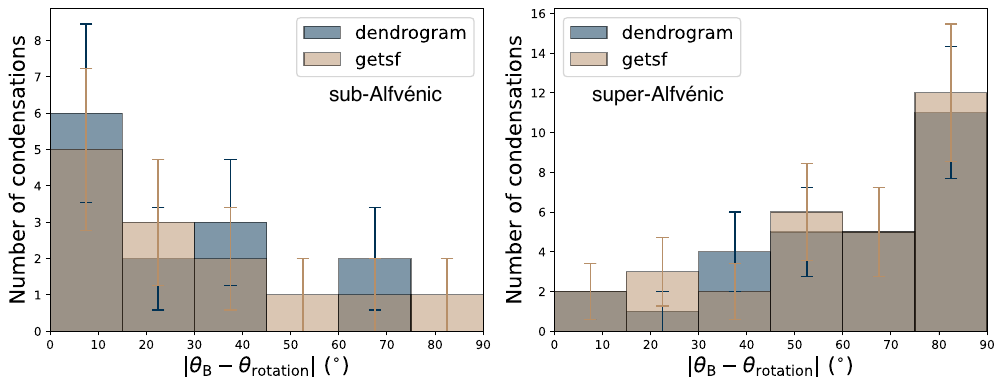}
\caption{\textbf{B-rotation alignment from simulations.} The histogram indicates the angular differences between the average B field orientation and the rotation axis of condensations for initially sub-Alfv\'{e}nic (left) and super-Alfv\'{e}nic (right) simulations, summed over 3 orthogonal planes. 
} \label{fig:sim_Brot}
\end{figure}

In initially sub-Alfv\'{e}nic simulations, the rotation axis exhibits a more parallel alignment with B fields (see Fig. \ref{fig:sim_Brot}, left), suggesting that a strong B field tends to cause the angular momentum to align with the fields \cite{2006ApJ...645.1227M}. 

In contrast, initially super-Alfv\'{e}nic models show a preferential misalignment between the rotation axis and B fields (see Fig. \ref{fig:sim_Brot}, right). In such turbulent conditions, the angular momentum of condensations is more likely to arise from local turbulent eddies than inherited from a coherent large-scale rotation \cite{2013A&A...554A..17J}. As condensations evolve, they tend to settle into a state of minimal energy and rotate around the shortest axis, the axis with the maximal moment of inertia \cite{2007JQSRT.106..225L}. Simultaneously, turbulence could align B fields with the elongation of condensations, naturally resulting in a misalignment between the B field and rotation axis. 

Note that our simulations are in ideal MHD conditions, while non-ideal MHD effects may also modify the B-rotation process \cite{2018FrASS...5...39W}. Thus, it is possible that some of the observed misalignment may be attributed to non-ideal MHD effects rather than turbulence.

Our observations provide statistical evidence for a preferential misalignment between magnetic (B) fields and the rotation axes of condensations. Supported by numerical investigations, we suggest that turbulence can plausibly produce this misalignment. Such misalignment has important implications on disk formation, as it can substantially reduce the efficiency of magnetic braking, thereby facilitating the formation of large and massive protostellar disks. This process can enable the formation of high-mass stars via disk-mediated accretion. Moreover, the observed misalignment conditions make disk fragmentation, a key mechanism for producing stellar multiplicity, more feasible than in aligned cases \cite{2013A&A...554A..17J}. 

\newpage
\section*{Methods}\label{sec:methods}

\subsection*{Observations and data reduction}

The observational data were obtained from the B fields in Massive Star-Forming Regions (MagMaR) survey. All the MagMaR targets have the potential to form high-mass stars \cite{2024ApJ...974...95I}. Polarization observations toward 30 MagMaR regions were carried out by ALMA between 2018 September 25 and 2021 May 4 under projects 2017.1.00101.S and 2018.1.00105.S (PI: Patricio Sanhueza) in Band 6 and configuration C-4. Three spectral windows were configured to observe the polarized dust continuum at $\sim$242.548-244.422 GHz, $\sim$244.548-246.422 GHz, and $\sim$256.5-258.4 GHz in the full polarization mode. The details of the observational setup can be found in previous MagMaR papers \cite{2021ApJ...915L..10S, 2021ApJ...923..204C, 2021ApJ...913...29F, 2024ApJ...972..115C, 2024ApJ...972L...6S, 2024ApJ...974..257Z, 2025ApJ...980...87S}. 

Data calibration was performed using the Common Astronomy Software Applications \citep[CASA;][]{2007ASPC..376..127M}. Three rounds of phase-only self-calibration were applied to the dust continuum data. Line emission channels were removed from the continuum following ref. \cite{2021ApJ...909..199O}. Imaging of the Stokes $I$, $Q$, and $U$ maps was carried out using the CASA task \texttt{tclean}. The synthesized beam major axes range from $\sim$0.27$''$ to $\sim$0.55$''$, corresponding to spatial resolutions of $\sim$450–1970 au at source distances of 1.3–5.3 kpc. The maximum recoverable scale (MRS) was $\sim$5$''$. A summary of the basic information of the MagMaR sources is provided in Table \ref{tab1}, including the coordinates, distance, 1$\sigma$ root-mean-square (RMS) noise level for Stokes $I$ maps ($\sigma_{I}$), and RMS for Stokes $Q$ and $U$ maps ($\sigma_{pol}$). Both $\sigma_{I}$ and $\sigma_{pol}$ are before primary beam correction. All the ALMA images shown in this paper are before primary beam correction. All the continuum fluxes used for calculations of density and mass are after primary beam correction. 

The debiased polarized intensity ($PI$) and its uncertainty ($\sigma_{PI}$) are calculated as $PI = \sqrt{Q^2 + U^2 - \sigma_{pol}^2}$ \cite{2006PASP..118.1340V} and $\sigma_{PI} \approx \sqrt{2}\sigma_{pol}$. The polarization position angle is computed using $\theta_{\mathrm{p}} = 0.5\arctan(U/Q)$, and its uncertainty is given by $\delta\theta = 0.5 \sqrt{\sigma_{pol}^2 / (Q^2 + U^2)}$ \cite{1993A&A...274..968N}. The B field orientation is inferred by rotating the polarization angle by 90 degrees. Here we assume the dust polarization is primarily due to the radiative torque alignment (RAT), as other possible dust polarization mechanisms are only predominant at scales smaller than 100–200 au \cite{2018ApJ...856L..27G}. 

Following ref. \cite{2018ApJ...861...14C}, we imaged the CH$_3$CN $14_3$–$13_3$ (257.4828 GHz) line, which falls within one of the configured spectral windows. The line RMS is approximately $\sigma_{line} \sim$0.5 mJy beam$^{-1}$ at 1.1 km s$^{-1}$ velocity intervals.

\subsection*{Simulations and synthetic observations}

We adopt 11 clustered massive star formation simulations of self-gravitating molecular clumps from ref.\cite{2021ApJ...919...79L}. The 3D ideal magneto-hydrodynamic (MHD) simulations were made with the RAMSES code \cite{2002A&A...385..337T}. The initial mass is 300 $M_{\odot}$ with a Plummer-like density profile within a 1- or 2-pc box. The initial B field is along the x-axis, and its strength is proportional to the column density. The initial magnetic level is controlled by the mass-to-flux ratio normalized to its critical value ($\mu$) \cite{1976ApJ...210..326M}. We include initial (super)sonic turbulence by seeding an initial turbulent velocity field with a random-phase Kolmogorov-like power spectrum \cite{2011A&A...528A..72H}. The initial temperature ($T$) is 10 or 20 K, and the initial turbulent level is controlled by the sonic Mach number ($\mathcal{M}$). To follow the gravitational collapse, we employ adaptive mesh refinement based on the local Jeans length \cite{1997ApJ...489L.179T}. The maximum refine level is 14, corresponding to a maximum resolution of 12.5 or 25 au. When the maximum level of refinement is reached, sink particles (i.e., protostars) are introduced \cite{2014MNRAS.445.4015B}. Among the models, T10M1MU1 and T10M3MU1 are initially sub-Alfv\'{e}nic (i.e., B fields dominate turbulence), while the remaining nine are initially super-Alfv\'{e}nic (i.e., turbulence dominates B fields). All the modeled clumps are sub-virial (i.e., gravity dominates turbulence) and magnetically super-critical (i.e., gravity dominates B fields), consistent with observations \cite{2013ApJ...779..185K, 2022ApJ...925...30L, 2024ApJ...966..120L}. The initial parameters of the numerical models are summarized in Table \ref{tab2}. We analyze the simulations at an intermediate evolutionary stage when the star formation efficiency (SFE) reaches 15\%. 

To model polarized dust emission, we adopt RAT as the primary grain alignment mechanism and post-process the MHD simulation outputs with the POLARIS radiative transfer code \cite{2016A&A...593A..87R}. This step includes recomputing the dust temperature and generating synthetic 1.3 mm Stokes $I$, $Q$, and $U$ maps from three orthogonal viewing directions ($xy$, $zy$, and $xz$ planes). The maps are convolved to a resolution of 1000 au. We test and verify that varying resolutions between 500 and 2000 au do not significantly impact the analysis results.  Given that the maximum recoverable scale of our ALMA observations is much larger than the condensation sizes, interferometric filtering is unlikely to affect the measured properties of condensations \cite{2016ApJ...820...38H, 2024ApJ...972..115C} and is therefore not considered in the synthetic observations. Gaussian noise, equivalent to the thermal noise levels in our ALMA data, is added to the synthetic maps to simulate observational conditions. The polarization position angle is calculated as $\theta_{\mathrm{p}} = 0.5 \arctan(U/Q)$, and the inferred B field orientation is obtained by rotating $\theta_{\mathrm{p}}$ by 90 degrees. The details of the radiative transfer modeling can be found in ref.\cite{2021ApJ...919...79L}.
%

Additionally, we produce synthetic spectral cubes of the CH$_3$CN $14_3 - 13_3$ (257.4828 GHz) line at the three viewing directions using POLARIS to trace dense gas kinematics. A CH$_3$CN abundance of $10^{-9}$ relative to H$_2$ is assumed \cite{2023ApJ...950...57T}, and line parameters are adopted from the LAMDA database \cite{2005AA...432..369S}. The spectral cubes are convolved to a resolution of 1000 au and injected with Gaussian noise to simulate observational conditions. 

Because our simulations do not include ionization feedback, they may not fully capture the physics of sources at very late evolutionary stages (e.g., those with HII regions). In addition, detailed disk accretion and outflow-driven processes are not included in our simulations. If a condensation has the same projected elongation direction as a disk, disk-driven outflows could align the magnetic field with the outflow cavity walls, potentially producing a perpendicular condensation–B alignment. Therefore, the preferentially parallel condensation–B alignment observed in our sample is unlikely to be primarily driven by outflow-related effects. 

\subsection*{Source identification}

We identify compact sources (i.e., condensations) in the ALMA dust continuum images using the dendrogram-based Python package {\tt astrodendro} \cite{2008ApJ...679.1338R}. This algorithm analyzes the hierarchical structure of emission by tracing the topology of isosurfaces across different intensity levels. The dendrogram construction requires three input parameters: the minimum intensity level for structure detection ($min_{-}value$), the minimum intensity contrast between different structures ($min_{-}delta$), and the minimum number of pixels for a structure to be considered ($min_{-}npix$). We set $min_{-}value$ = 5$\sigma_{I}$, $min_{-}delta$ = 1$\sigma_{I}$, and $min_{-}npix$ to half the number of pixels within the synthesized beam area. The {\tt astrodendro} method outputs the coordinates, full width at half maximum (FWHM) along the major and minor axes, and position angle ($\theta_{\mathrm{condensation}}$) for each identified condensation (leaf). Within the elliptical area of each condensation, we compute the average B field orientation by averaging the Stokes $Q$ and $U$ values of pixels with $PI/\sigma_{PI} > 3$. A total of 898 condensations are identified across the 30 regions, of which 480 show polarization detections greater than half the beam area. Most condensations have FWHM sizes ranging from several hundred au to approximately 2000–3000 au. A few outliers with unrealistically large FWHMs are excluded from our analysis. Using similar parameter settings, we also apply {\tt astrodendro} to identify condensations in the synthetic dust continuum maps from the numerical models, projected along three orthogonal planes ($xy$, $zy$, and $xz$). Examples of dendrogram-identified condensations in both observations and simulations are shown in Extended Data Figs. 1 and 2. Sources with size $>$0.05 pc are not used in our analysis. 


To ensure that our analysis is not significantly biased by the specific source identification and fitting algorithm used in {\tt astrodendro}, we also employ the {\tt getsf} \cite{2021A&A...649A..89M} method to identify the condensations. The {\tt getsf} is a method for extracting sources and filaments in astronomical images using the separation of their structural components. In this work, we only focus on the identified compact sources (condensations). The method requires a single input parameter: the maximum size of sources to extract. We adopt 5$''$, approximately ten times the synthesized beam size. The {\tt getsf} method outputs the coordinates, FWHM along the major and minor axes, and position angle for each identified condensation.  A total of 883 condensations are identified, of which 424 show polarization detections greater than half the beam area. The same procedure is applied to synthetic dust emission maps from the numerical simulations to identify the condensations. Examples of identified condensations are shown in Extended Data Figs. 1 and 2. The B field orientation for each condensation is computed from the averaged polarization signal within its area.

\subsection*{Velocity gradient of condensations}
To characterize the internal kinematics of condensations, we analyze the velocity centroid ($v_c$) maps of the CH$_3$CN $14_3 - 13_3$ line, a common tracer of hot molecular cores (HMCs). With a high critical density ($n_\mathrm{crit} = 4.5 \times 10^6$ cm$^{-3}$) and upper energy level ($E_\mathrm{u}/k_{\mathrm{B}} = 157$ K), this transition is well-suited for probing dense and warm gas associated with ongoing massive star formation. 

Among the 30 observed regions, CH$_3$CN emission is detected toward 182 condensations identified in dust continuum by {\tt astrodendro} and 163 by {\tt getsf}. The relatively small fraction of condensations exhibiting CH$_3$CN line emission suggests that most condensations are not in the HMC phase. For each region, we generate velocity centroid (intensity-weighted) maps by selecting velocity channels with signal-to-noise ratios greater than $5\sigma_\mathrm{line}$. 

To quantify the velocity structure within individual condensations, we fit the $v_c$ distribution inside  the elliptical region as defined by {\tt astrodendro} and {\tt getsf} using a linear velocity gradient model \cite{1993ApJ...406..528G}:
\begin{equation}
v_c = v_0 + a\Delta\alpha + b\Delta\beta,
\end{equation}
where $v_0$ is the systemic velocity, $\Delta\alpha$ and $\Delta\beta$ are offsets in right ascension and declination, and $a$ and $b$ represent the projected velocity gradients along those axes. The orientation of the velocity gradient, $\theta_{\mathrm{vg}}$, is computed as $\theta_{\mathrm{vg}} = \tan^{-1}(a/b)$.

The quality of the fit is assessed using the coefficient of determination ($R^2$), which ranges from 0 to 1, with higher values indicating better fits. For the subsequent analysis, we select only those condensations with $R^2 > 0.8$, which also visually exhibit a clear linear velocity gradient. This yields 62 condensations from {\tt astrodendro} and 59 from {\tt getsf}. 

We define rotation-like condensations as those where the velocity gradient direction is more aligned (offsets $<$30$^{\circ}$) with the major axis of the condensation \cite{1997ApJ...475..211O}. We test and verify that varying the offset limit (but still $<$45$^{\circ}$) does not significantly impact the analysis results. Note that infall motions may still exist, but do not dominate in those rotation-like condensations. The orientation of the rotation axis ($\theta_{\mathrm{rotation}}$) is inferred by rotating $\theta_{\mathrm{vg}}$ by 90 degrees. A similar procedure is applied to the synthetic CH$_3$CN data from the numerical simulations to derive the rotation axis of simulated condensations.



\backmatter

\vspace{1em}
\noindent\textbf{Data Availability}

This paper makes use of the following ALMA data: ADS/JAO.ALMA\#2017.1.00101.S and ADS/JAO.ALMA\#2018.1.00105.S. The data are available at \url{https://almascience.nao.ac.jp/aq} by setting the observation code. The reduced ALMA data and the simulation data used for this study are available from the corresponding author upon reasonable request. The catalog of condensation parameters in the ALMA dust continuum maps, the POLARIS output synthetic images, and an example of the RAMSES output are available under https://doi.org/10.5281/zenodo.19062258.


\vspace{0.2em}
\noindent\textbf{Code Availability}

The ALMA data were reduced using CASA version 6.5.5-21, which is available at \url{https://casa.nrao.edu/casa_obtaining.shtml}. The source identification package of {\tt astrodendro} is available at \url{http://dendrograms.org/}.  The source identification package of {\tt getsf} is available at \url{https://irfu.cea.fr/Pisp/alexander.menshchikov/}. This research made use of Astropy, a community-developed core Python package for Astronomy \cite{2013A&A...558A..33A}, and Matplotlib, a Python 2D plotting library for Python \cite{2007CSE.....9...90H}.


\vspace{0.2em}
\noindent\textbf{Acknowledgements}

J.L. thanks Dr. Anaelle Maury for her constructive comments on this work. J.L. thanks Dr. Pak Shing Li, Prof. Kenichi Tatematsu, and Prof. Hua-Bai Li for helpful discussions. 
J.L. was partially supported by Grant-in-Aid for Scientific Research (KAKENHI Number 23H01221 and 25K17445) of the Japan Society for the Promotion of Science (JSPS). 
P.S. was partially supported by a Grant-in-Aid for Scientific Research (KAKENHI Number JP22H01271 and JP23H01221).
P.S. was partially supported by a Grant-in-Aid for Scientific Research (KAKENHI Nos. JP22H01271 and JP24K17100) of the JSPS. 
J.M.G. acknowledges the support from the program Unidad de Excelencia María de Maeztu CEX2020-001058-M and the grant PID2020-117710GB-I00 (MCI-AEI-FEDER, UE). 
X.L. acknowledges support from the National Key R\&D Program of China (No. 2022YFA1603101), the Strategic Priority Research Program of the Chinese Academy of Sciences (CAS) Grant No. XDB0800300, the National Natural Science Foundation of China (NSFC) through grant Nos. 12273090 and 12322305, the Natural Science Foundation of Shanghai (No. 23ZR1482100), the CAS “Light of West China” Program No. xbzg-zdsys-202212, and 100101 Key Laboratory of Radio Astronomy and Technology (Chinese Academy of Sciences). 
K.Q. acknowledges support from National Natural Science Foundation of China (NSFC) grants 12425304 and U1731237, and National Key R\&D Program of China No. 2023YFA1608204 and No. 2022YFA1603100.
S.L. acknowledges support from the National SKA Program of China with No. 2025SKA0140100, “Double First-Class” Funding with No. 14912217, and National Natural Science Foundation of China (NSFC) grant with No. 13004007.
M.T.B. acknowledges financial support through the INAF Large Grant  {\it The role of magnetic fields in massive star formation} (MAGMA).
This paper makes use of the following ALMA data: ADS/JAO.ALMA\#2017.1.00101.S and ADS/JAO.ALMA\#2018.1.00105.S. ALMA is a partnership of the ESO (representing its member states), NSF (USA), and NINS (Japan), together with NRC (Canada), MOST and ASIAA (Taiwan), and KASI (Republic of Korea), in cooperation with the Republic of Chile. The Joint ALMA Observatory is operated by ESO, AUI/NRAO, and NAOJ. The National Radio Astronomy Observatory is a facility of the National Science Foundation operated under cooperative agreement by Associated Universities, Inc. Data analysis was in part carried out on the Multi-wavelength Data Analysis System operated by the Astronomy Data Center (ADC), National Astronomical Observatory of Japan.


\vspace{0.2em}
\noindent\textbf{Author contributions}

J.~L led the simulations, data analysis, interpretation of results, and paper writing. P.~Sanhueza led the ALMA proposal. P.~Saha and P.~Sanhueza contributed to the ALMA data imaging. P.~C and J.~M.~G assisted with the ALMA data imaging. K.~M contributed to the dendrogram analysis of ALMA data. S.~C and J.~K contributed to the hot molecular core identification. B.~C contributed to the RAMSES simulations. V.~V assisted with the POLARIS simulations. All authors discussed the results and commented on the manuscript. 


\vspace{0.2em}
\noindent\textbf{Competing interests}

The authors declare no competing interests.

\clearpage

\setcounter{figure}{0}
\renewcommand{\figurename}{Extended Data Fig.}
\renewcommand{\figureautorefname}{Extended Data Fig.}

\begin{figure}[htb]
\centering
\includegraphics[width=0.95\textwidth]{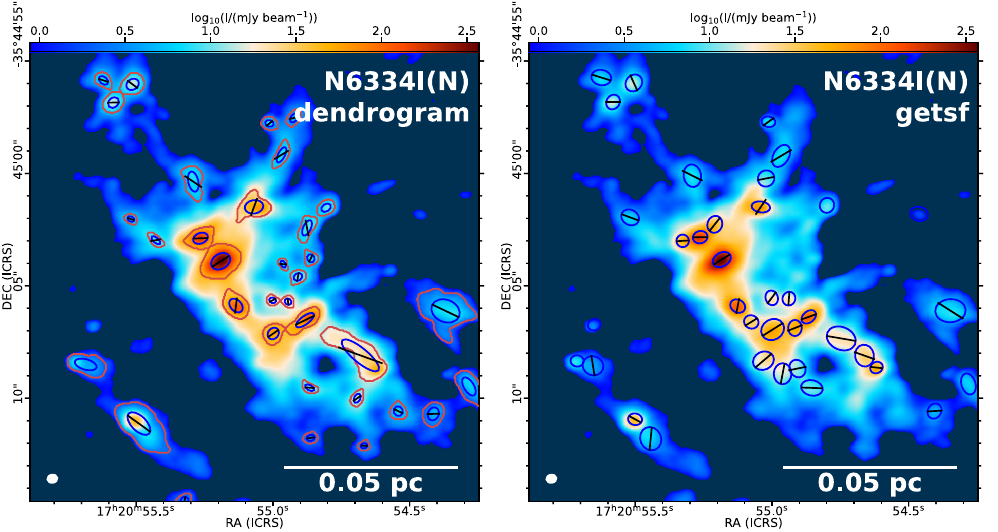}
\caption{\textbf{Dust intensity structures of observed massive protoclusters.} Examples of ALMA 1.3 mm dust continuum maps. The blue ellipses indicate the condensations identified with the FWHM along the major and minor axis and the position angle, as reported by {\tt astrodendro} (left) and {\tt getsf} (right). The black lines indicate the average B field orientation of each condensation. In the left panel, the red contours indicate the mask of the condensations identified by {\tt astrodendro}. 
} \label{fig:obs_dendro}
\end{figure}

\begin{figure}[htb]
\centering
\includegraphics[width=0.95\textwidth]{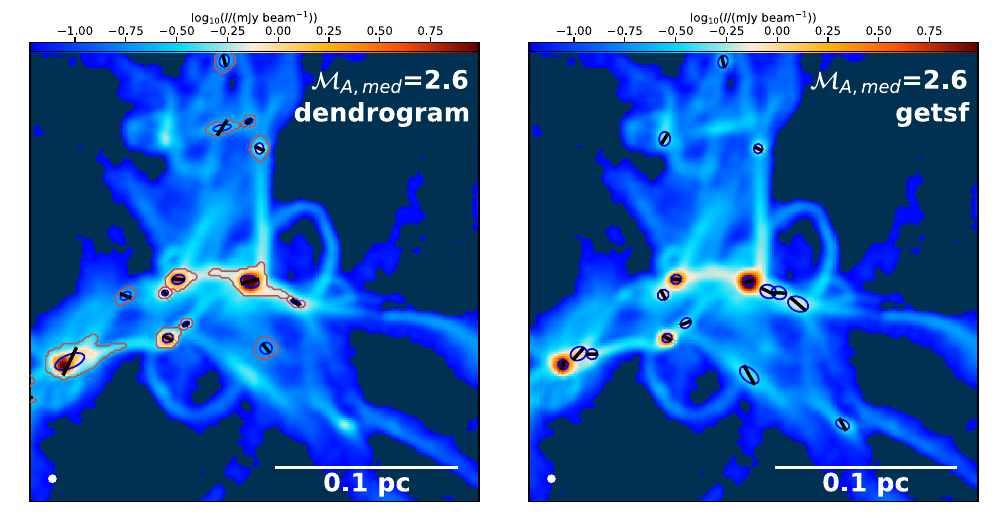}
\caption{\textbf{Dust intensity structures of simulated massive protoclusters.} Examples of synthetic 1.3 mm dust continuum maps (T10M6MU2). The blue ellipses indicate the condensations identified by {\tt astrodendro} (left) and {\tt getsf} (right). The black lines indicate the average B field orientation of each condensation. In the left panel, the red contours indicate the mask of the condensations identified by {\tt astrodendro}. 
} \label{fig:sim_dendro}
\end{figure}

\begin{figure}[htb]
\centering
\includegraphics[width=0.95\textwidth]{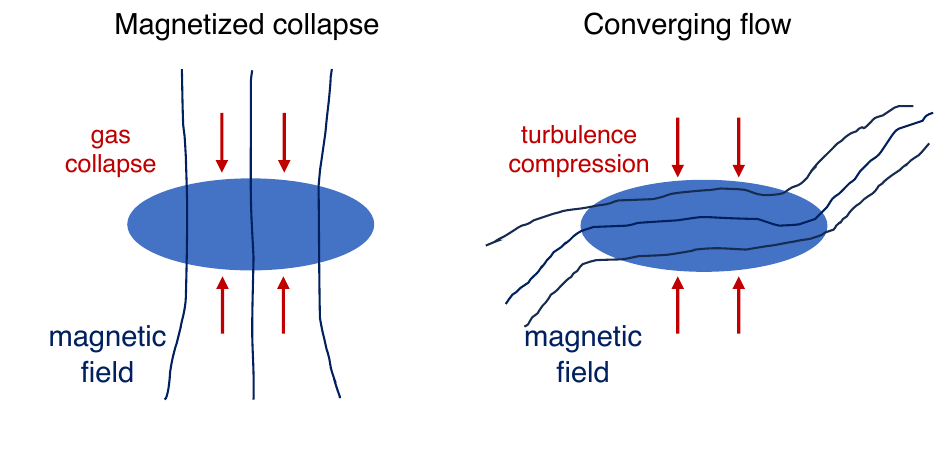}
\caption{\textbf{Schematic view of dense structure formation.} Left: Dense structure form by gravo-magnetic processes. Right: Dense structure form by gravo-turbulent processes. (This figure is added after publish of the paper and is thus not peer-reviewed).
} \label{fig:schem}
\end{figure}

\begin{appendices}



\clearpage

\setcounter{figure}{0}
\renewcommand{\figurename}{Supplementary Fig.}
\renewcommand{\figureautorefname}{Supplementary Fig.}

\section*{Supplementary Information}
\subsection*{Information of observations and simulations}\label{App:obsandsim}

The basic information of the observed sources is summarized in Supplementary Table \ref{suptab1}. The initial parameters of the numerical models are summarized in Supplementary Table \ref{suptab2}. 

\begin{table}[!htb]
\caption{\textbf{Information of observed sources from the ALMA survey.} The coordinates, distance, and sensitivity of each source are cataloged.}\label{suptab1}%
\centering
\begin{tabular}{@{}llllll@{}} 
\toprule
Source & RA  & DEC & Distance & $\sigma_{I}$ & $\sigma_{pol}$ \\ 
& (h:m:s) & (d:m:s) & (kpc) & ($\mu$Jy beam$^{-1})$ & ($\mu$Jy beam$^{-1})$ \\ 
\midrule
     G5.89 & 18:00:30.43 & -24:04:01.64 &  2.99 &   450 &    38\\
    G10.62 & 18:10:28.65 & -19:55:49.52 &  4.95 &    69 &    39\\
     G11.1 & 18:10:28.27 & -19:22:30.92 &   3.0 &    50 &    29\\
    G11.92 & 18:13:58.02 & -18:54:19.02 &  3.37 &   181 &    30\\
   G14.22S & 18:18:13.00 & -16:57:21.82 &   1.9 &    66 &    12\\
     G24.6 & 18:35:40.50 &  -7:18:34.02 &  3.45 &    47 &    23\\
    G29.96 & 18:46:03.76 &  -2:39:22.60 &  5.26 &    39 &    27\\
 G34.43MM1 & 18:53:18.01 &   1:25:25.50 &   3.5 &   210 &    23\\
 G34.43MM2 & 18:53:18.58 &   1:24:45.98 &   3.5 &   120 &    20\\
   G35.03A & 18:54:00.65 &   2:01:19.30 &  2.32 &    94 &    22\\
    G35.13 & 18:58:06.30 &   1:37:05.98 &   2.2 &   123 &    22\\
    G35.2N & 18:58:13.03 &   1:40:36.00 &  2.19 &   233 &    21\\
   G333.12 & 16:21:36.00 & -50:40:50.01 &   3.3 &   115 &    30\\
   G333.23 & 16:19:51.20 & -50:15:13.00 &   5.2 &    97 &    29\\
   G333.46 & 16:21:20.26 & -50:09:46.56 &   2.9 &   160 &    29\\
  G335.579 & 16:30:58.76 & -48:43:54.01 &  3.25 &   149 &    32\\
   G335.78 & 16:29:47.00 & -48:15:52.32 &   3.2 &   170 &    32\\
   G336.01 & 16:35:09.30 & -48:46:48.16 &   3.1 &   137 &    31\\
   G351.77 & 17:26:42.53 & -36:09:17.40 &   1.3 &  1300 &    58\\
    I16547 & 16:58:17.22 & -42:52:07.49 &   2.9 &   424 &    23\\
    I16562 & 16:59:41.63 & -40:03:43.62 &  2.38 &   470 &    25\\
    I18089 & 18:11:51.40 & -17:31:28.52 &  2.34 &   175 &    30\\
    I18151 & 18:17:58.17 & -12:07:25.02 &   3.0 &    62 &    19\\
    I18162 & 18:19:12.20 & -20:47:29.02 &   1.3 &   154 &    32\\
    I18182 & 18:21:09.13 & -14:31:50.58 &  3.58 &   118 &    19\\
    I18223 & 18:25:08.55 & -12:45:23.32 &   3.4 &    44 &    18\\
    I18337 & 18:36:40.82 &  -7:39:17.74 &   3.8 &    57 &    18\\
    N6334I & 17:20:53.30 & -35:47:00.02 &  1.35 &  2029 &   121\\
 N6334I(N) & 17:20:54.90 & -35:45:10.02 &  1.35 &   439 &    51\\
      W33A & 18:14:39.40 & -17:52:01.02 &  2.53 &   131 &    30\\
\bottomrule
\end{tabular}
\end{table}
%

\begin{table}[!htb]
\caption{\textbf{Initial parameters of numerical models.} The box size, maximum resolution, temperature, sonic Mach number, normalized mass-to-flux ratio, and median Alfv\'{e}n Mach number of each model are cataloged.}\label{suptab2}%
\centering
\begin{tabular}{@{}lllllll@{}}
\toprule
Model & box & resolution & $T$  & $\mathcal{M}$ & $\mu$ & $\mathcal{M}_{\mathrm{A,med}}$\footnotemark[1]\\
& (pc) & (au) & (K) & & &\\
\midrule
T10M1MU1 & 2 & 25 & 10 & 1 & 1.2 & 0.23\\
T10M3MU1 & 2 & 25 & 10 & 3 & 1.2 & 0.70\\
T10M3MU20 & 2 & 25 & 10 & 3 & 20 & 11\\
T10M6MU2 & 2 & 25 & 10 & 6.4 & 2 & 2.6 \\
T10M6MU200 & 2 & 25 & 10 & 6.4 & 200 & 261 \\
T20M3MU2 & 1 & 12.5 & 20 & 3 & 2 & 1.3  \\
T20M3MU5 & 1 & 12.5 & 20 & 3 & 5 & 2.9  \\
T20M3MU200 & 1 & 12.5 & 20 & 3 & 200 & 116  \\
T20M6MU2 & 1 & 12.5  & 20 & 6.4 & 2 & 2.6  \\
T20M6MU5 & 1 & 12.5  & 20 & 6.4 & 5 & 6.5  \\
T20M6MU200 & 1 & 12.5 & 20 & 6.4 & 200 & 261 \\
\bottomrule
\end{tabular}
\footnotetext[1]{Initial median Alfv\'{e}n Mach number, which is the ratio between the non-thermal velocity dispersion of the whole simulation and the Alfv\'{e}n velocity of the median B field value.}
\end{table}

\subsection*{Source identification parameter}\label{App:para}
Here we explore how the input paramters of {\tt astrodendro} and {\tt getsf} could affect the source identification results and the corresponding condensation-B alignment trends. 

For {\tt astrodendro}, we vary each input parameter ($min_{-}value$, $min_{-}delta$, and $min_{-}npix$) by a factor of 2, except for $min_{-}value$, where the the lower limit is set to 3$\sigma_{I}$ for scientifically meaningful analysis. Supplementary Fig. 1 shows the CDFs of condensation-B alignment for different parameter combinations. Overall, the resulting condensation-B alignment trend is not sensitive to different combinations of input parameters.  

\begin{figure}[htb]
\centering
\includegraphics[width=0.45\textwidth]{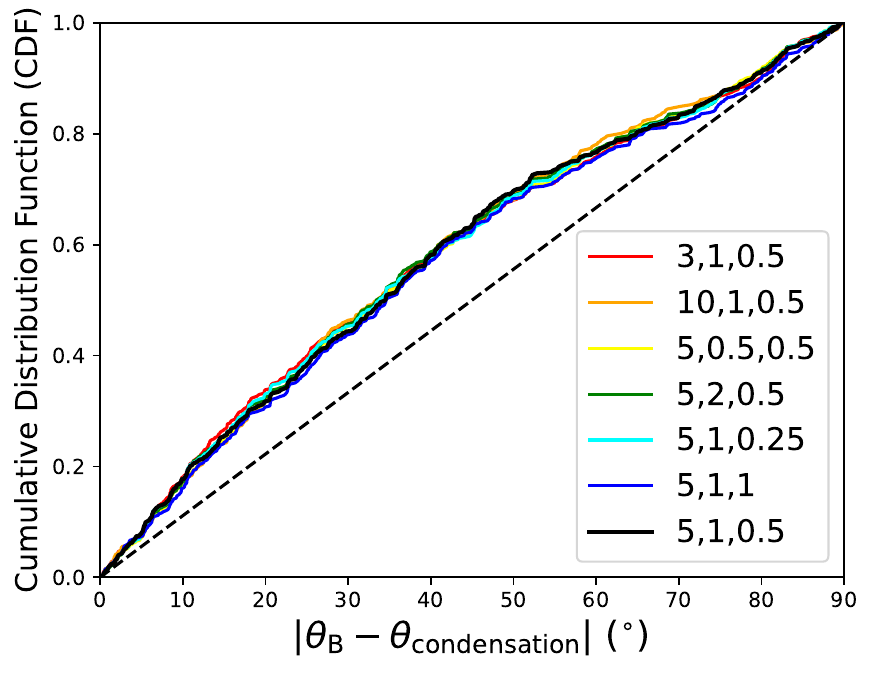}
\caption{\textbf{Cumulative distribution function of condensation-B alignment.} The dashed black line represents a random alignment distribution. The solid black line shows the CDF from our adopted parameter combination. Colored solid lines represent CDFs from different parameter combinations ($min_{-}value$ in units of $\sigma_{I}$, $min_{-}delta$ in units of $\sigma_{I}$, $min_{-}npix$ in units of beam area). }\label{fig:obs_dendro_cdf}
\end{figure}

For {\tt getsf}, we vary the maximum source size by a factor of 2 and find that the identified condensations and the resulting condensation-B alignment do not change. This is reasonable because the identified condensations are much smaller than our choice of maximum source size, and the identification of sources using {\tt getsf} is not sensitive to this parameter \cite{2021A&A...649A..89M}.

\subsection*{Projection effect}\label{App:proj}
For any two vectors in three-dimensional (3D) space, the angle between them as projected onto the plane of the sky ($\phi_{\mathrm{POS}}$) can differ substantially from their true 3D angle ($\phi_{\mathrm{3D}}$). For instance, two vectors that are more misaligned in 3D can appear more aligned in 2D projection, and vice versa \cite{2014ApJ...792..116Z}. Thus, we perform Monte Carlo simulations to assess how this projection effect may impact our statistical results.

We begin by generating 1000 pairs of randomly oriented vectors uniformly distributed in 3D space. Note that the uniformly distributed quantity in 3D is the trigonometric function $\cos \phi_{\mathrm{3D}}$ (see Supplementary Fig. 2a) instead of the angle $\phi_{\mathrm{3D}}$ \cite{2013ApJ...774..128S}. Thus, we characterize the alignment in 3D with $\cos \phi_{\mathrm{3D}}$. Then we project the 3D intersecting angles onto the plane of the sky at a fixed viewing direction. The resulting distribution of the projected angles ($\phi_{\mathrm{POS}}$) for these vector pairs is shown in Supplementary Fig. 2b. The randomly distributed vectors in 3D leads to a flat distribution of $\phi_{\mathrm{POS}}$ from 0$^{\circ}$ to 90$^{\circ}$. 

Next, we consider the subset of vector pairs that are intrinsically more aligned in 3D, defined by $\cos \phi_{\mathrm{3D}} > 0.5$ (i.e., $\phi_{\mathrm{3D}} < 60^\circ$), which accounts for roughly half of the full sample. The distribution of their projected angles still shows a preferential alignment, although projection causes approximately 29\% of these intrinsically aligned pairs to appear misaligned ($\phi_{\mathrm{POS}} > 45^\circ$) in 2D (Supplementary Fig. 2b). 

We repeat this analysis for the other half of the sample with $\cos \phi_{\mathrm{3D}} < 0.5$ (i.e., $\phi_{\mathrm{3D}} > 60^\circ$), corresponding to intrinsically more misaligned vector pairs. In this case, projection causes approximately 31\% of the intrinsically misaligned pairs to appear more aligned ($\phi_{\mathrm{POS}} < 45^\circ$), yet the projected angle distribution still shows an overall trend toward misalignment (Supplementary Fig. 2b).

The angle distributions in the Monte Carlo simulations are idealized, which may not reflect the true condensation-B alignment. Thus, we have additionally explored the projection effect with our numerical data. Similar to the 2D condensation identification, we identify 3D condensations in the simulated density cubes using {\tt astrodendro}. The density data are convolved with a Gaussian kernel of 1000 au to match the observational resolution. For the parameter of {\tt astrodendro}, we adopt $min_{-}value = 10^6$ cm$^{-3}$, approximately corresponding to the lower density limit of condensations \cite{2022ApJ...925...30L}. The $min_{-}delta$ is set to $0.05 \times min_{-}value$ in logarithmic scale, and $min_{-}npix$ is chosen as half the number of pixels within the 3D Gaussian beam volume. 

For each condensation identified by the dendrogram masks, we fit the density distribution to derive the density-weighted principal axes, approximating each condensation as a 3D ellipsoid. We compute the volume-averaged B field orientation and measure the 3D angular difference ($\phi_{\mathrm{3D}}$) between the B field and the longest principal axis of each condensation. We then project each ellipsoid onto three orthogonal planes ($xy$, $yz$, and $xz$), and measure the corresponding 2D condensation–B angular difference ($\phi_{\mathrm{2D}}$). We find that the trend of projected condensation-B alignment is overall consistent with the intrinsic 3D alignment trend, although a minority of condensations exhibit a change in alignment after projection (see Supplementary Fig. 2c). This result is similar to those found in the Monte Carlo simulations. 

\begin{figure}[htb]
\centering
\includegraphics[width=0.95\textwidth]{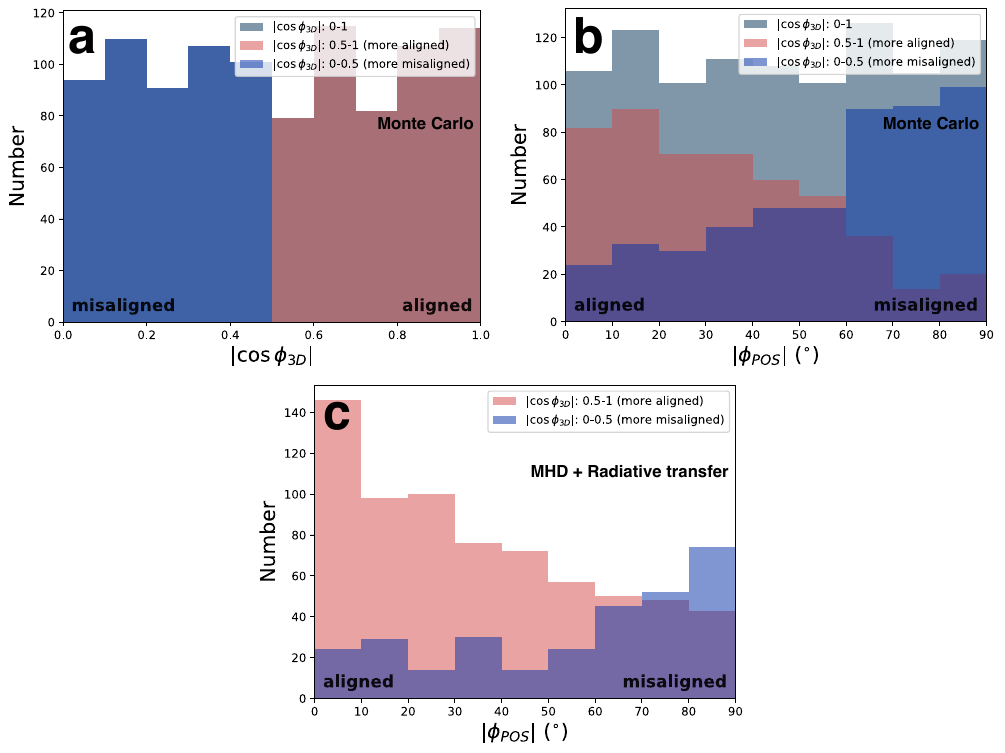}
\caption{\textbf{Distribution of alignment between vector pairs.} \textbf{a}, Histogram of the 3D alignment (quantified by $\cos \phi_{\mathrm{3D}}$) for vector pairs from Monte Carlo simulations in three groups: uniformly distributed (gray), intrinsically more aligned with $\cos \phi_{\mathrm{3D}} > 0.5$ (red), and more misaligned with $\cos \phi_{\mathrm{3D}} < 0.5$ (blue). \textbf{b}, Histogram of the corresponding projected angles $\phi_{\mathrm{POS}}$ for the same three groups of vector pairs from Monte Carlo simulations. \textbf{c}, Histogram of the projected angular difference $\phi_{\mathrm{POS}}$ between B field orientation and condensation elongation for two intrinsic alignment regimes ($\cos \phi_{\mathrm{3D}} > 0.5$ and $\cos \phi_{\mathrm{3D}} < 0.5$) derived from our numerical simulations.
} \label{fig:mc_angle}
\end{figure}

Our tests show that although the projection effect can significantly influence individual measurements, the general trends in the projected angle distributions still can reflect the underlying 3D alignment. This provides confidence that our observed trends are not solely artifacts of projection but reflect true physical alignment in 3D.

\subsection*{Condensation aspect ratio}\label{App:roundness}

The elongation of more circular condensations is less well defined, which can introduce uncertainties when comparing their orientations with the B field \cite{2009MNRAS.399.1681T}. To evaluate this effect, we investigate the correlation between condensation roundness and the angular offset between condensation elongation and the B field direction, based on our ALMA observations.

Supplementary Fig. 3a shows the mean angular difference between condensation elongation and the B field orientation in different aspect ratio bins, where each bin contains an equal number of condensations. The aspect ratio is calculated as the ratio of the major to minor FWHM, as reported by {\tt astrodendro} and {\tt getsf}. For nearly circular condensations (aspect ratio $\sim$1), the mean angular difference approaches the value expected for a random distribution (i.e., 45$^\circ$). As the aspect ratio increases (more elongated), the B field orientation becomes more aligned with the elongation of the condensations. Overall, the uncertainty from the roundness of condensations makes the general alignment trend less clear, but does not strengthen or reverse it. Condensations identified by {\tt astrodendro} generally exhibit larger aspect ratios than those identified by {\tt getsf}, which is likely because {\tt getsf} separates compact sources from elongated background structures before fitting the FWHM of sources. 

As cores and disks are preferentially oblate-like structures at larger and smaller scales \cite{2007MNRAS.379L..50T}, it is reasonable to assume that most condensations are also oblate-like rather than prolate-like or irregular in 3D. In this context, the observed trend is naturally explained if the 3D B field is preferentially misaligned with the shortest axis of the oblate \cite{2009MNRAS.399.1681T}: when an oblate condensation is viewed face-on, its projected elongation appears more circular, making the condensation-B offset more uncertain and closer to a random distribution. When viewed edge-on, the projected elongation is more clearly defined, and alignment with the B field becomes more evident.


\begin{figure}[htb]
\centering
\includegraphics[width=0.95\textwidth]{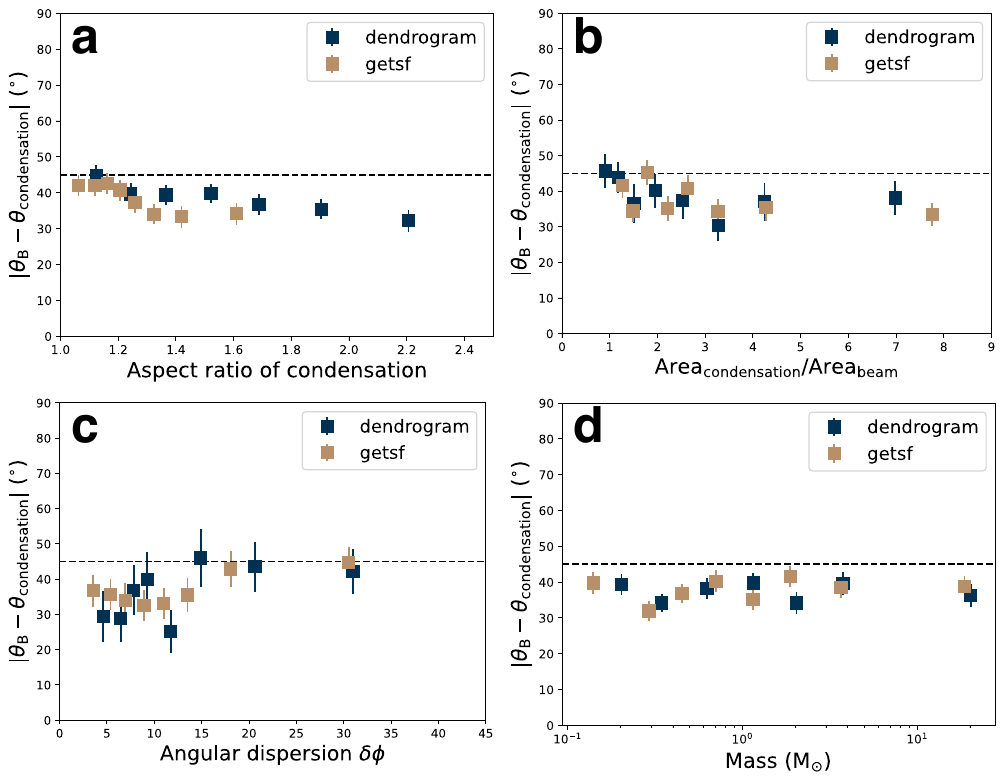}
\caption{\textbf{Relation between condensation-B alignment and different parameters.} The mean angular offset between condensation elongation and average B orientation from ALMA observations is shown in bins of (a) condensation aspect ratio; (b) condensation-beam area ratio; (c) angular dispersion of condensation B field; (d) condensation mass. Error bars represent the standard error of mean. The dashed line marks an angular difference of 45$^\circ$, corresponding to a random distribution.
} \label{fig:align}
\end{figure}

\subsection*{Beam smoothing effect}\label{App:beam}

Due to the finite resolution of observations, the emission profiles of condensations are smoothed, which can introduce uncertainties in the estimation of the condensation elongation. To assess this effect, we examine how beam smoothing influences the condensation-B alignment in our ALMA data.

Supplementary Fig. 3b shows the mean condensation-B angular offset in bins of area ratio between condensations and the ALMA synthesized beam, with each bin containing an equal number of condensations. For smaller condensations, the mean offset approaches the value expected from a random distribution, while for larger condensations, the B field and condensation elongation appear more aligned. This trend indicates that beam smoothing increases the uncertainty in condensation-B offsets for marginally resolved condensations, whereas the alignment distribution of well-resolved condensations is less affected. Overall, the uncertainty from the beam smoothing effect makes the general alignment trend less clear, but does not strengthen or reverse the trend.

\subsection*{Dispersion of B field orientation} \label{App:angdis}

The dispersion of B field orientation within each condensation brings uncertainties in the estimation of the condensation-averaged field orientation. Here we investigate how the condensation-B alignment is affected by the angular dispersion of B in the ALMA data.

Supplementary Fig. 3c shows the mean condensation-B angular offset in bins of B field angular dispersion ($\delta \phi$), with each bin containing an equal number of condensations. The calculation of $\delta \phi$ is performed with circular statistics. For condensations with large $\delta \phi$, the mean offset approaches the value expected from a random distribution, whereas for condensations with small $\delta \phi$, the B field and condensation elongation appear more aligned. This trend indicates that for condensations with complex B structures, the average field orientation is less well defined, leading to larger uncertainties in the condensation-B comparison. In contrast, condensations with more uniform B-field patterns show more reliable alignment measurements. 

It is important to note that the relation between the Alfv\'{e}n Mach number $\mathcal{M}_{\mathrm{A}}$ and the angular dispersion $\delta \phi$ is given by $\mathcal{M}_{A} \sim (f_t/f_u/f_o/Q_c) \delta\phi_{\mathrm{obs}}$ under the assumptions of the Davis–Chandrasekhar–Fermi (DCF) method, where $f_t$ accounts for the 3D to POS conversion of the turbulent B field, $f_u$ accounts for the 3D to POS conversion of the ordered B field, $f_o$ is a factor to account for the ordered field contribution to the angular dispersion, and $Q_c$ is a factor to correct for other uncertainties on the assumptions in the calculation \cite{2022FrASS...9.3556L}. Therefore, a larger $\delta \phi$ does not necessarily imply a higher $\mathcal{M}_{\mathrm{A}}$ (i.e., more turbulent) if the various correction factors are not properly taken into account \cite{2021ApJ...919...79L}. 

\subsection*{Condensation mass and emission} \label{App:align_M}

Following ref. \cite{2024ApJ...974...95I}, we estimated the mass of the observed condensations by assuming a constant temperature and dust emissivity index within each region. The temperature of the parental structures derived in ref. \cite{2024ApJ...974...95I} is adopted as the condensation temperature, with the caveat that condensation masses may be overestimated if the actual condensation temperature exceeds that of the parental structure. Here we examine the relation between condensation–B alignment and condensation mass in the ALMA data.

Supplementary Fig. 3d shows the mean condensation–B angular offset in bins of condensation mass, with each bin containing an equal number of condensations. Overall, no strong correlation is found between the condensation-B alignment and condensation mass. Similarly, we find no significant correlations between condensation–B alignment and either peak intensity or total flux of the condensation. This means that the condensation–B alignment has no strong dependence on the condensation mass, and that the observed alignment trend is not significantly affected by the signal-to-noise ratio of the identified condensations. 

Supplementary Fig. 4 shows the mass-radius relation for the condensations identified in the ALMA dust continuum maps. We adopt the FWHM from 2D Gaussian fitting (by {\tt astrodendro} and {\tt getsf}) as the radius of the condensations. About 65\% ({\tt astrodendro}) and 45\% ({\tt getsf})) of the condensations satisfy the theoretical surface density threshold of 1 g cm$^{-2}$ for massive star formation \cite{2008A&A...487..993K}. 

\begin{figure}[htb]
\centering
\includegraphics[width=0.95\textwidth]{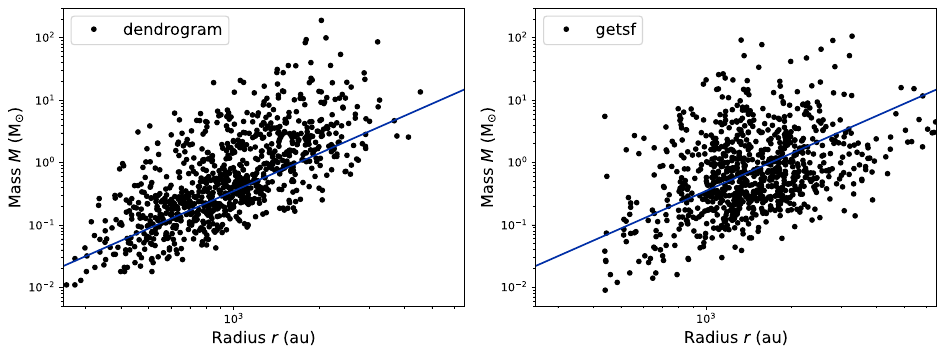}
\caption{\textbf{Mass-radius relation for observed condensations.} Mass vs. radius for condensations identified in the ALMA dust continuum maps. The blue line indicates the theoretical surface density threshold of 1 g cm$^{-2}$ for massive star formation \cite{2008A&A...487..993K}.
} \label{fig:obs_gleaf}
\end{figure}

\subsection*{Condensation-gravity alignment} \label{App:gleaf}

Gravitational flows along B field lines may form elongated condensed structures along the flowing path, which could also result in parallel alignment between B fields and condensation elongations \cite{2023ApJ...943...76M}. To assess this effect, we investigate the correlations between the condensation elongation and the gravity direction in each observed region. 

Following Ref. \cite{2012ApJ...747...79K}, we have calculated the mapwise 2D gravity direction ($\theta_{\mathrm{G}}$) of the observed regions from the mass distribution maps using the standard formula of gravity. The mean direction of gravity within each identified condensation is derived using circular statistics. Supplementary Fig. 5 compares the mean gravity direction with the condensation elongation for all the condensations. Overall, no strong correlation is found between the two orientations. This means that gravitational contraction alone cannot explain the observed condensation-B alignment. 

\begin{figure}[htb]
\centering
\includegraphics[width=0.45\textwidth]{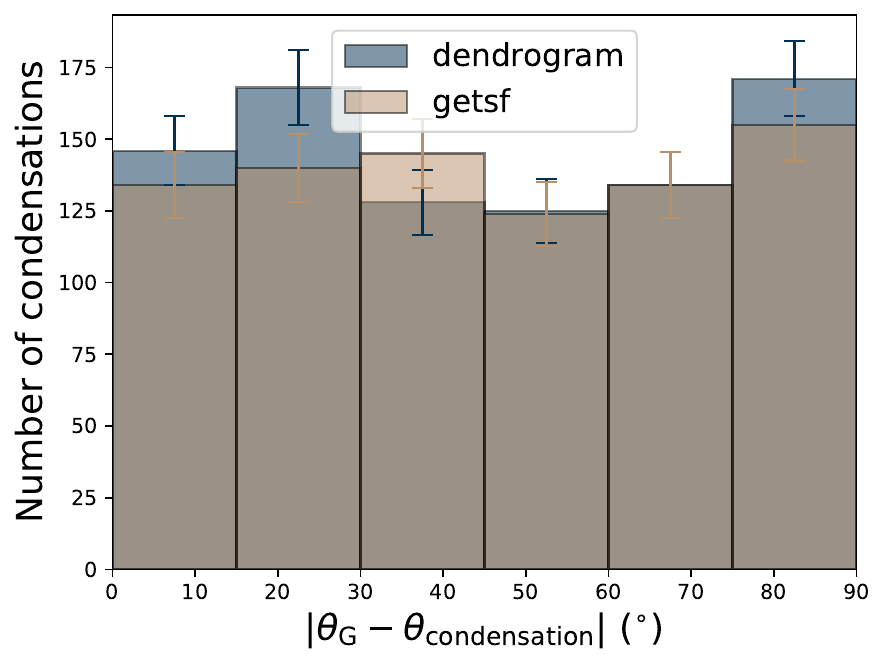}
\caption{\textbf{Condensation-gravity alignment for observed condensations.} Histograms of the angular difference between condensation elongation and gravity direction across all observed regions.
} \label{fig:obs_gleaf}
\end{figure}

\subsection*{Tracing rotation with line velocity gradient} \label{App:rot}

It has been analytically suggested that a velocity gradient along the major axis of an isolated elliptical core primarily traces rotational motions \cite{1997ApJ...475..211O}. In clustered regions, however, the velocity field is more complex. Thus, we perform an analysis with our numerical simulations to test whether velocity gradients in such environments can still trace the rotational motions of condensations.

In the investigation of the projection effect in a previous Supplementary subsection, we have identified 3D condensations in the simulation data and approximated each condensation as a 3D ellipsoid. Within each ellipsoid, the direction of the total angular momentum is adopted as the intrinsic 3D rotation axis. We then project each ellipsoid onto three orthogonal planes ($xy$, $yz$, and $xz$). Similar to the approaches applied to ALMA data, we fit the velocity gradient within the projected elliptical area using synthetic CH$_3$CN line data, and infer the 2D rotation axis for condensations where the velocity gradient is well fitted and closely aligned with the major axis.

Supplementary Fig. 6 compares the projected intrinsic rotation axis with the axis inferred from CH$_3$CN line fitting. The fitted rotation axis is preferentially aligned with the actual projected rotation axis, supporting the validity of using CH$_3$CN velocity gradients to trace the rotation axis of condensations. 

\begin{figure}[htb]
\centering
\includegraphics[width=0.45\textwidth]{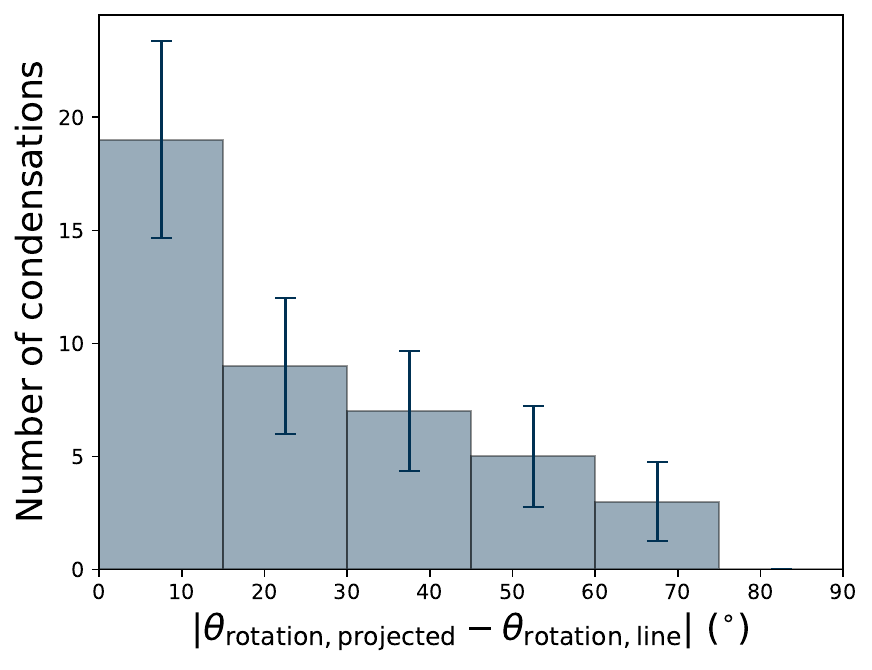}
\caption{\textbf{Comparison between intrinsic and fitted rotation axes in simulations.} Histogram of angular offsets between the projected intrinsic rotation axis of condensations and the rotation axis inferred from fitting CH$_3$CN velocity gradients.}
\label{fig:sim_rot_proj}
\end{figure}
\end{appendices}

\bibliography{sn-article}

\end{document}